\documentclass[aps,prl,twocolumn,groupedaddress]{revtex4-2}

\usepackage{amsmath,amssymb,mathrsfs,graphicx,comment}
\usepackage{dcolumn}
\usepackage{bm}
\usepackage{color}

\begin{document}
\title{Phase-resolved attosecond near-threshold photoionization of molecular nitrogen}
\author{S. Haessler$^1$, B. Fabre$^2$, J. Higuet$^2$, J. Caillat$^3$, T. Ruchon$^1$, P. Breger$^1$, B. Carr\'e$^1$, E. Constant$^2$, A. Maquet$^3$, E. M\'evel$^2$, P. Sali\`eres$^1$,  R. Ta\"{\i}eb$^3$ and Y. Mairesse$^2$ }
\affiliation{$^1$CEA-Saclay, IRAMIS, Service des Photons, Atomes et Mol\'ecules, 91191 Gif-sur-Yvette, France} 
\affiliation{$^2$CELIA, Universit\'e Bordeaux I, UMR 5107 (CNRS, Bordeaux 1, CEA), 351 Cours de la Lib\'eration, 33405 Talence Cedex, France}
\affiliation{$^3$UPMC, Universit\'e Paris 06, CNRS, UMR 7614, LCPMR, 11 rue Pierre et Marie Curie, 75231 Paris Cedex 05, France}

\begin{abstract}
{\color{blue}This is a corrected version of the manuscript published as [Phys. Rev. A \textbf{80}, 011404(R) (2009)].}
We photoionize nitrogen molecules with a train of extreme ultraviolet attosecond pulses together with a weak infrared field. We measure the phase of the two-color two-photon ionization transition (molecular phase) for different states of the ion. We observe a 0.76 $\pi$ shift for the electrons produced in the ionization channels leading to the X$^2 \Sigma_g^+$, $v'=1$ and $v'=2$ states. We relate this phase shift to the presence of a complex resonance in the continuum. By providing both a high spectral and temporal resolution, this general approach gives access to the evolution of extremely short lived states, which is hardly accessible otherwise. 
\end{abstract}
\pacs{}
\maketitle

Ionization of atoms and molecules by absorption of extreme ultraviolet (XUV) radiation provides rich structural information on the considered species. The ionization process releases an electron wavepacket, which can be described as a coherent superposition of partial waves. The relative contributions and phases of the partial waves can be extracted from photoelectron angular distributions, {\em at a given energy} \cite{Lebech03}. However, the temporal structure of the ejected wavepacket, which is imposed by the phase relation between different energy components, is not accessible with such experiments. To access this phase, one needs to couple two energy components of the electron wavepacket and record the resulting interference. This can be achieved by absorption of high order harmonics of an infrared laser pulse, in the presence of the fundamental field.

\begin{figure}[t]
\begin{center}
\includegraphics[width=.4\textwidth]{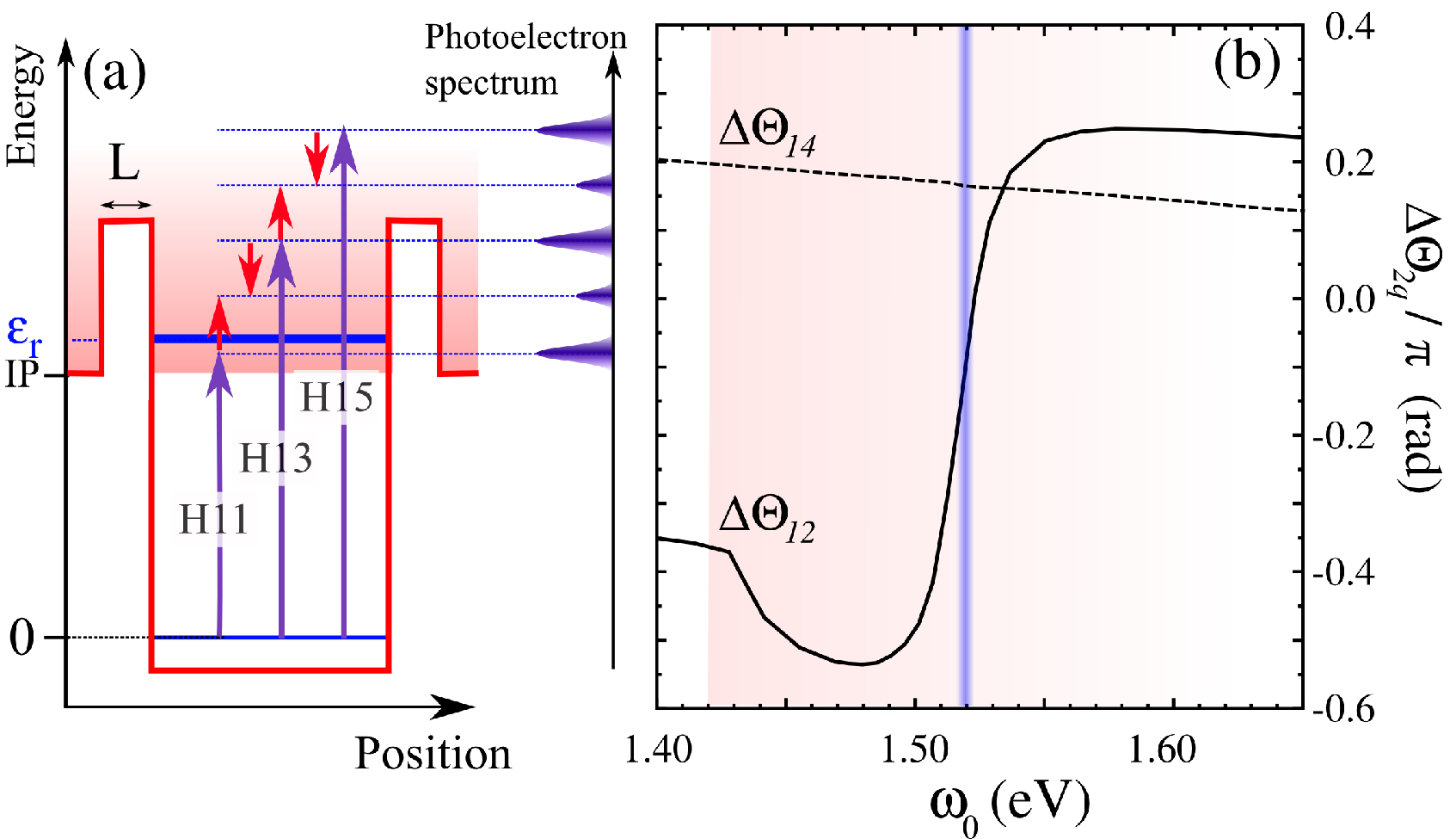}
\end{center}
\caption{\label{FigTheory} Theoretical model of a RABBITT experiment in presence of a resonance. 
(a) Principle of RABBITT and model potential that is used for the simulations. (b) 
Evolution of the molecular phase difference $\Delta\theta_{12}$ (solid line) and $\Delta\theta_{14}$ (dashed line) as a function of the fundamental laser frequency. Harmonic 11 is in the continuum for $\omega_0 > 1.43$ eV and hits the resonance when $\omega_0 = 1.52$ eV.
}
\end{figure}

An intense laser pulse propagating in a gas jet produces coherent XUV radiation constituted of odd harmonics $(2q+1)\omega_0$ of the fundamental frequency $\omega_0$. These harmonics are all approximately phase locked with the fundamental and form an Attosecond Pulse Train (APT) \cite{Paul01}. In photoionization experiments with high harmonics, the photoelectron spectrum exhibits equidistant lines resulting from single photon ionization (Fig. \ref{FigTheory}(a)). If an additional laser field with frequency $\omega_0$ is added, two-photon ionization can occur: absorption of a harmonic photon accompanied by either absorption or stimulated emission of one photon $\omega_0$. New lines (sidebands) appear in the spectrum, in between the harmonics (Fig. \ref{FigTheory}(a)). Since two coherent quantum paths lead to the same sideband, interferences occur. They are observed in an oscillation of the sideband amplitude as the delay $\tau$ between the probe (IR) and harmonic fields is scanned \cite{Veniard96,Paul01}. This is the basis of the RABBITT technique (Reconstruction of Attosecond Beating By Interference of Two-photon Transitions). The phase of the oscillation is determined by the phase difference between consecutive harmonics (phase locking) and by additional phases characteristic of the ionization process.

The same process can be described in the time domain. The APT creates a train of attosecond electron wavepackets. The additional laser field acts as an optical gate on the electrons, which can be used to retrieve the temporal profile of the electron wavepackets \cite{Quere05,MairessePRA05}. This temporal structure is set by the temporal shape of the APT but also by the photoionization process. Thus, performing RABBITT measurements with a well characterized APT gives access to the spectral phase of the photoionization \cite{Varju05,Mauritsson05}, i.e. the temporal dynamics of photoionization. 

Recently Cavalieri \textit{et al.} reported a time-resolved measurement of photoionization of a solid target by a single attosecond pulse \cite{Cavalieri07}. Conceptually this is close to RABBITT \cite{Quere05,MairessePRA05} but using of an APT rather than a single pulse has major advantages: (i) the production of APT is much less demanding ; (ii) the spectrum of APT is a comb of narrow harmonics that can be used to identify different photoionization channels ; (iii) the intensity of the IR beam must be of  $\approx 10^{11}$ W.cm$^{-2}$ for RABBITT and about $10^{13}$ W.cm$^{-2}$ with single pulses \cite{Kienberger04}, which can perturb the system. 

Here we study the photoionization of nitrogen molecules with an APT and characterize the outgoing electron wavepackets using the RABBITT technique. We probe the region just above the ionization threshold of N$_2$, which is spectroscopically very rich \cite{Parr81,Raoult83,Zubek88}. We show that the ``complex resonance'' (at 72.3 nm) \cite{Raoult83,Dehmer84,Zubek88} induces a $\approx \pi$ phase change in the molecular phase. This effect strongly depends on the final ionic state, revealing the complex nature of the photoionization process.

We first present a theoretical study of the influence of resonances on RABBITT measurements. The oscillation of sideband  $S_{2q}$ between two harmonics $2q-1$ and $2q+1$ is given by \cite{Veniard96}:
\begin{equation}
S_{2q}\propto \cos(2\omega_0\tau+\varphi_{2q+1}-\varphi_{2q-1}+\theta_{2q+1}-\theta_{2q-1}),
\end{equation}
where $\varphi_{2q\pm1}$ are the phases of the harmonics H$({2q\pm1})$. 

The quantities $\theta_{2q\pm1}$ are the phases of second order transition matrix  $T_{2q}^{(\pm)}$ corresponding to the absorption of harmonics $2q\pm1$ accompanied by either the emission (-) or absorption (+) of one IR photon, both paths leading to $S_{2q}$. In the perturbative regime, we have
\begin{eqnarray}\label{secord}
T_{2q}^{(\pm)}&=&\sum_{n}\hspace{-.5cm}\int\hspace{.2cm}\langle k_{2q}|{\cal D}|n\rangle\langle n|{\cal D}|0\rangle \nonumber \\
&\times&\left[\frac{1}{-I_p+(2q\pm1)\omega_0-\epsilon_n}+\frac{1}{-I_p\mp\omega_0-\epsilon_n}\right].
\end{eqnarray}
where ${\cal D}$ is the dipole operator, $|0\rangle$ the initial state and $|k_{2q}\rangle$ the final continuum states labeled by the momentum of the outgoing electron ($k_{2q}^2/2=\epsilon_0+2q\omega_0$). The sum is running over all states (discrete and continuum) $ |n\rangle$ of energy $\epsilon_n$ of N$_2$. 

The aim of our work is the measurement of the quantities $\theta_{2q\pm1}$, which are generally referred to as atomic or molecular phases \cite{Toma02,Mauritsson05,Varju05}. In structureless continua, i.e., rare gases and for high enough harmonics, the difference $\Delta\theta_{2q}=\theta_{2q+1}-\theta_{2q-1}$ tends to vanish. However, close to the ionization threshold or in the presence of a resonance, this is no longer the case. As we will show, RABBITT permits to measure such molecular phases and to uncover their variations in the vicinity of a resonance.

To this end, we have simulated a RABBITT experiment by solving the time dependent Schr\"odinger equation for an electron in a 1-D model potential interacting with a perturbative IR pulse and several of its odd harmonics  (Fig. \ref{FigTheory}(a)).
The potential shape was adjusted to get the same ionization energy as N$_2$ (Ip=15.58 eV) and an autoionizing resonant state with energy $\epsilon_r=16.7$ eV which can be reached by the 11$^{th}$ harmonic of a 800-nm laser pulse. The resonance width $\Gamma_r$ can be tuned by adjusting the barrier width $L$. We show in Fig.\ref{FigTheory}(b) the computed phase differences  as a function of $\omega_0$ for a barrier width $L= 6$ a.u. corresponding to a lifetime $\Gamma_r^{-1}\approx57 $fs.

While $\Delta\theta_{14}$ slightly varies with the IR frequency, $\Delta\theta_{12}$ is strongly dependent on the detuning of H11 relative to the resonance energy. When H11 is below the resonance, i.e. $\omega_0 < 1.52$ eV, there is an important phase difference between $\Delta\theta_{12}$ and $\Delta\theta_{14}$,  while when H11 is above the resonance, $\Delta\theta_{12}$ is close to $\Delta\theta_{14}$. 
This can be understood with Eq.(\ref{secord}): in the first component of the sum, the term corresponding to the resonant state becomes dominant when $\-Ip+11\omega_0\approx\epsilon_r$, as its denominator changes sign. In the limit of an infinite resonance lifetime ($\Gamma_r \rightarrow 0$), this would lead to a $\pi$-jump in the molecular phase. With a finite lifetime, $\Gamma_r$ enters the denominator as the imaginary part of the resonance energy, smoothing the phase variation and reducing the magnitude of the jump.

The measurements were performed with the setup described in \cite{Mairesse03}. We used a 20-Hz 50-mJ 50-fs laser system. The laser beam is split into a generating beam and a delayed probe beam. The annular generating beam is focused into a pulsed argon jet and generates an APT. The generating IR is filtered out and the XUV radiation is focused by a grazing incidence Au-coated toroidal mirror in a molecular nitrogen effusive jet. The photoelectrons are detected with a magnetic bottle time-of-flight spectrometer. The probe beam is focused together with the XUV in the nitrogen jet and the IR-XUV delay is adjusted with a piezoelectric transducer. 

\begin{figure}[t]
\begin{center}
\includegraphics[width=.45\textwidth]{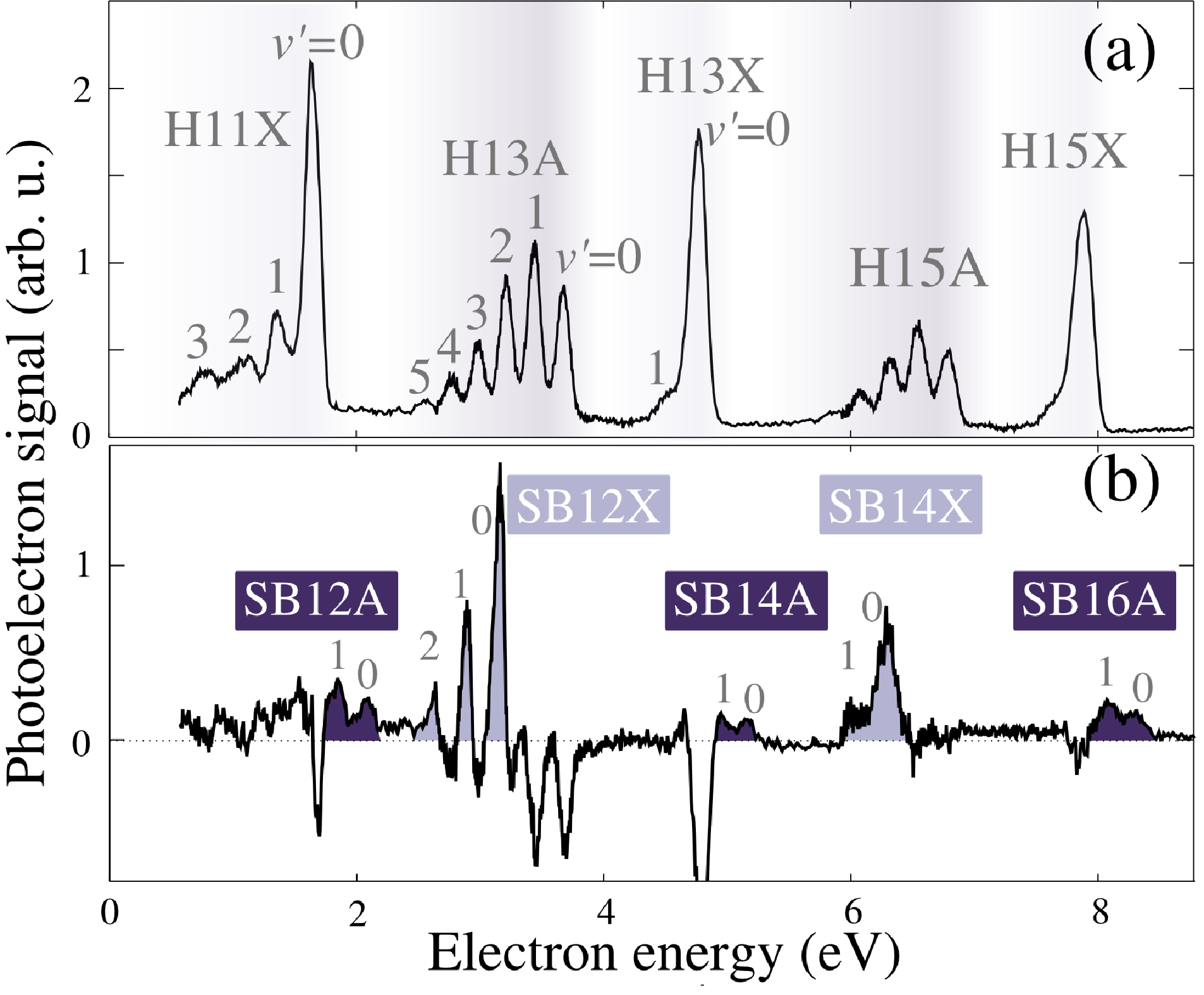}
\end{center}
\caption{\label{FigSpectres} Photoionization of N$_2$ by a comb of high-order harmonics and influence of the IR field. (a) Spectrum obtained with harmonics alone. (b) Difference between the photoelectron spectra obtained with and without the IR probe field.  }
\end{figure}

Figure \ref{FigSpectres}(a) shows a photoelectron spectrum obtained without the probe field. The energy resolution of the  spectrometer is maximal in the range 1-3 eV. We have applied different voltages to the flight tube so that the electrons produced by different harmonics would be detected in this optimal energy range. We concatenate the spectra to get the high resolution spectrum shown in Fig. \ref{FigSpectres}(a). Each spectrum was averaged over 2000 laser shots.

The harmonics are separated by twice the fundamental frequency, i.e. 3.13 eV. Each harmonic produces two photoelectron bands, corresponding to two ionization channels: one correlated to the fundamental electronic state X $^2\Sigma_g^+$ of the N$_2^{+}$ ion (Ip=15.58 eV), and the other to the excited A $^2\Pi_u$ state (Ip=16.69 eV). For each harmonic, the integral of the A band is about 1.8 times larger than that of the X band, in agreement with measured photoionization cross sections \cite{Itikawa86}. 

The photoelectron bands show fine structures revealing the vibrational states of the ion. The A and X states of N$_2^+$ have different equilibrium distances \cite{Lofthus77} and the corresponding bands present different shapes. For the ion ground state X, in which the equilibrium distance (1.12 \AA) is about the same as for the molecular ground state (1.10 \AA), we observe a narrow vibrational distribution. By contrast, for the first excited state of the ion A, the larger equilibrium distance (1.17 \AA) results in a broader vibrational distribution. While there is an overall good agreement between the observed vibrational populations and the Franck-Condon factors, the X band produced by H11 runs over more vibrational levels than expected. 

The spectrum of the fundamental laser pulse was centered at 1.565 eV, leading to a central energy of 17.22 eV for H11 with a spectral width of about 0.15 eV. The photoionization spectrum of N$_2$ shows a resonance within this  width, around 17.12 eV  \cite{Dehmer84}, which was attributed to the B $^2\Sigma_u^+$(3d$\sigma_g$) $^1\Sigma_u^+$  Hopfield Rydberg state \cite{Raoult83,Zubek88}. It is of particular interest because it is a ''complex resonance``, i.e. it involves a coupling with the highest members of another Rydberg series converging to the  A $^2\Pi_u$ $v^+\ge2$ \cite{Raoult83,Dehmer84}. 

Such quasibound autoionizing states are known to significantly modify the vibrational repartition, transferring part of the $v'=0$ population to higher vibrational states as $v'=1$, $2$ and $3$ \cite{Parr81}. Assuming that the states of the Hopfield series have the same shape as the B, we calculated the Franck-Condon factors for a transition from the molecular ground state to the different vibrational levels of the X state of the ion, via the first two vibrational levels of a B-like autoionizing state. We obtain the following populations for $v'=0,1,2,3,4$: 58\%, 26\%, 12\%, 4\%. This is in fair agreement with the measured populations : 58\%, 20\%, 12\%, 10\%. 

When the IR probe field is added, sidebands appear between the harmonics. The sidebands from one ionization channel partly overlap with the harmonics from the other ionization channel. In order to unambiguously identify them, we subtract the spectra obtained with and without the probe field (Fig. \ref{FigSpectres}(b)). The resulting signal is negative where the IR field has removed some population, i.e. where harmonics are located, and positive where it has increased the signal, i.e. where sidebands are. We label each sideband according to the corresponding photon energy and ion state. 

We scanned the delay between XUV and IR and measured the phase of the sideband oscillations by fast-Fourier-transform. This operation was performed for each sideband and each ionization channel. Different sidebands were measured with different retarding voltages, here again to achieve a correct resolution of the vibrational levels. In order to extract the molecular phase from the phase of the sideband oscillation (Eq. 1), one has to remove the difference $\varphi_{2q+1}-\varphi_{2q-1}$. This difference is inherent to the harmonic generation process and is independent of the target gas. It was measured by RABBITT of the same APT in a reference atom (argon) whose atomic phases are known from theory \cite{Mairesse03}. The term $\varphi_{2q+1}-\varphi_{2q-1}$ was then subtracted from the sideband phase to obtain $\Delta\theta_{2q}$.

\begin{figure}[t]
\begin{center}
\includegraphics[width=.5\textwidth]{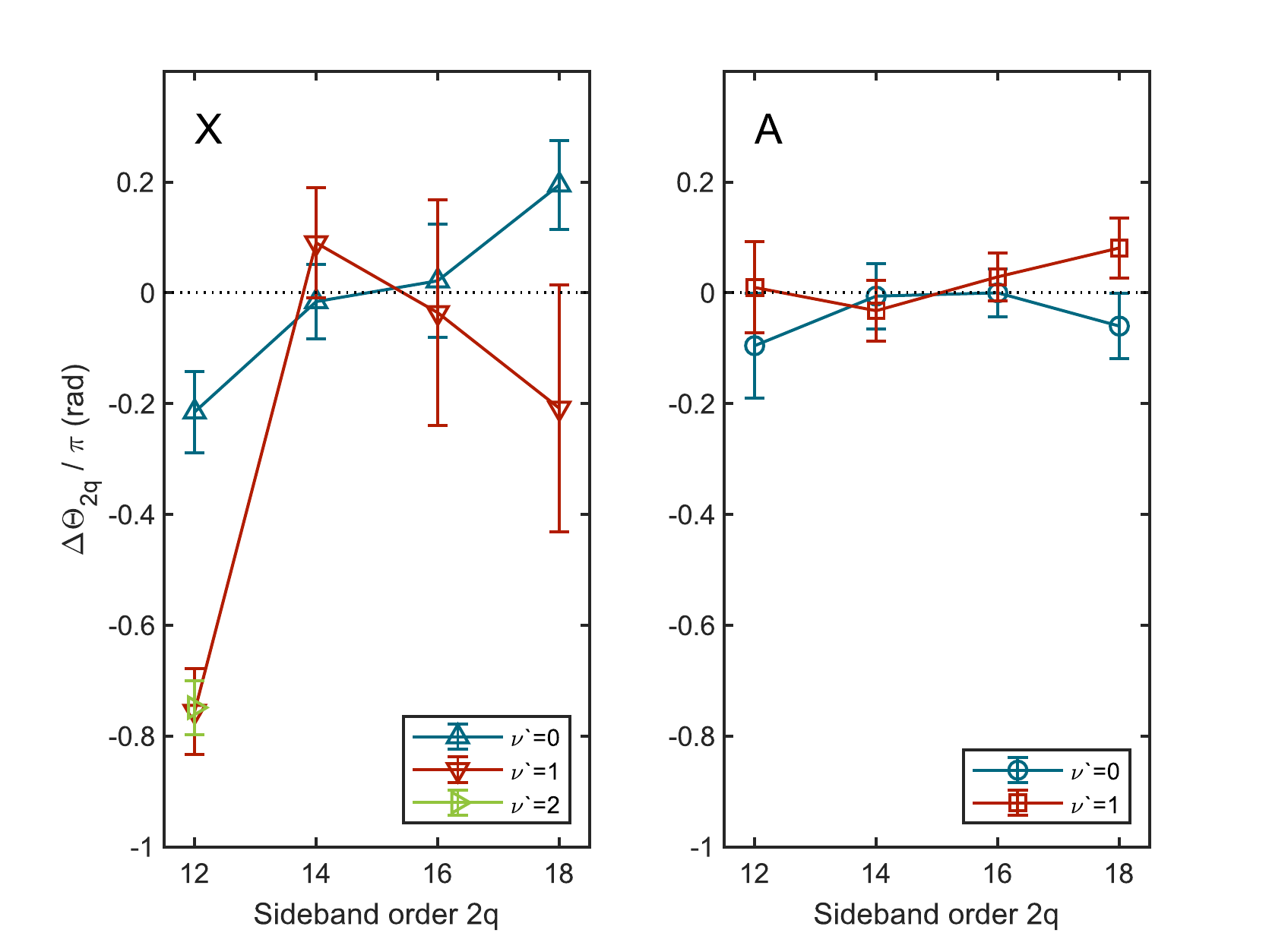}
\end{center}
\caption{\label{FigPhases} Molecular phase difference $\Delta\theta_{2q}$ as a function of the sideband order, for different ionization channels.}
\end{figure}

The evolution of $\Delta\theta_{2q}$ as a function of sideband order is shown in Fig. \ref{FigPhases}.
For the X channel, sideband 12 shows a remarkable behavior: $\Delta\theta_{12}$ is $-0.21 \pi$ for $v'=0$ and reaches $-0.76 \pi$ for $v'=1$ and $v'=2$.
This variation of $\Delta\theta_{12}$ is due to a phase jump in the two-photon matrix element associated to the absorption of H11. This phase shift was found to be very robust, even by varying the intensity of the IR probe field by a factor 3.
For the A channel, $\Delta\theta_{2q}$ stays close to zero for all sideband orders, between $-0.1\pi$ and $0.08\pi$. 
The absence of signature of the resonance in this channel is due to a poor interaction of the autoionizing state with the A continuum \cite{Raoult83}.

\begin{figure}[t]
\begin{center}
\includegraphics[width=.4\textwidth]{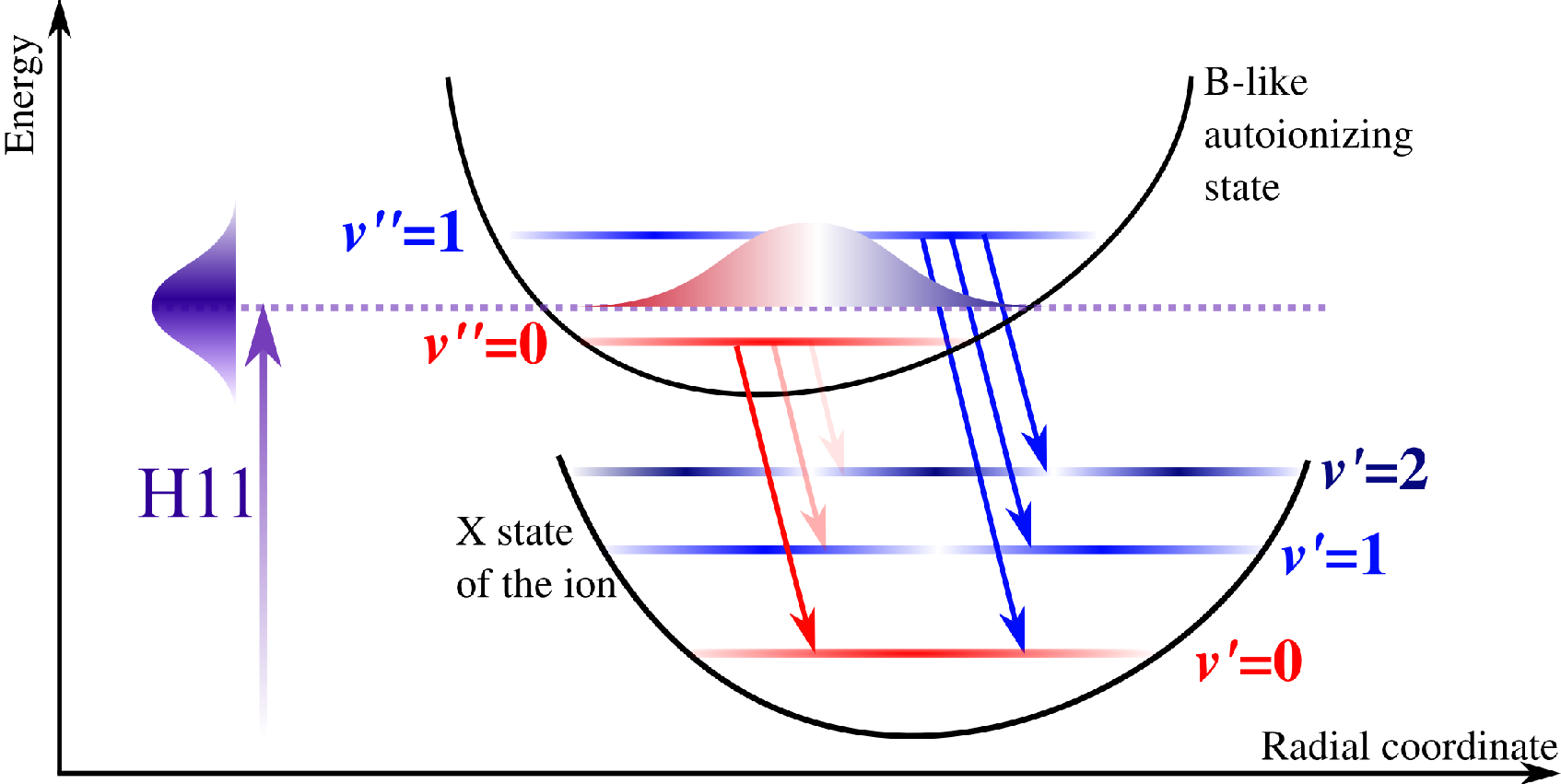}
\end{center}
\caption{\label{FigExplication} Schematic description of the photoionization process via the vibrational levels of the autoionizing state.  }
\end{figure}

While the signature of the vibrationally dependent influence of the resonance is already present in the amplitude distribution of the vibrational states (Fig. \ref{FigSpectres}), these measurements bring new information on the photoionization process. Indeed, as shown in Fig. \ref{FigTheory}, the effect of the resonance on the molecular phase is strongest when the energy of the harmonic is below the resonance. We can use this information to schematically represent the photoionization process (Fig. \ref{FigExplication}). The initial wavepacket is projected onto the first two vibrational levels ($v''=0,1$) of the B-like resonance that are contained within the bandwidth of H11. The central energy reached by absorption of H11 (horizontal dotted line in Fig. \ref{FigExplication}) is above the $v''=0$ level, so that the molecular phase is weakly affected by the resonance for this path. By contrast, H11 is below the $v''=1$ level and the molecular phase will present a $\approx\pi$ phase shift for this path. Franck-Condon factors show that the population from the $v''=0$ of a B-like state will mainly end in the $v'=0$ of the X ionic state, while the overlap of the $v''=1$  is good with the $v'=0,1,2$ states. Thus, for the X,$v'=0$ state, the signal is a mixture of direct and resonant contributions from the $v''=0$, for which there is no phase shift, and contributions from the $v''=1$ which are shifted. The resulting phase is only slightly affected by the resonance. By contrast, for the X,$v'=1$ and X,$v'=2$ states, the contributions from the $v''=1$ level dominate and there is a strong phase shift. The simple picture presented in Fig. \ref{FigExplication} is schematic since it does not explain why the phase shift for the $v'=1$ and $v'=2$ are equal. This is probably a signature of the coupling to Rydberg states converging to the A $^2\Pi_u$ $v^+\ge2$. 

Our results show that RABBITT is very sensitive to the location of the one-photon ionization level with respect to the vibrational states and to the shape of the autoionizing state potential energy surface. Thus, it can be used to reveal subtle information on the phase of the XUV photoionization process. The observed behavior of the two-photon molecular phases for different sidebands can be interpreted as a modification of the electron release time by the presence of the resonance. In the single XUV photoionization process, this would correspond to a change of the electron wavepacket temporal profile, admitting that the IR probe beam does not significantly perturb this process and does not induce an additional phase shift. While this assumption is consistent with our measurements of molecular phases independent of the probe beam intensity, it remains to be formally established.

In conclusion, we have used interference of two-photon transitions to measure the amplitude and phase of electron wavepackets produced by attosecond XUV photonionization of N$_2$ molecules. We have shown the influence of a complex resonance on the ionization phase for different vibrational states of the ion. 
Natural extensions of our work, with objective to uncover the exact nature of the resonances, would be (i) to measure the photoelectron angular distributions \cite{Aseyev03}; (ii) tuning the harmonic wavelength by tuning the IR wavelength \cite{MairessePRL05}, to scan the resonance. Further, the extension to the study of dynamics is quite straightforward and would enable to track the modifications of the photoionization phase in real-time in an excited medium. 

Last, we note that our measurement is conceptually close to coherent control experiments in which phase lags due to resonances can be measured through the interference of two paths leading to the same final state \cite{Fiss99,Yamazaki07}. The use of combined IR ﬁelds and APT opens the possibility to control molecular photoionization processes and to perform attosecond XUV coherent control.

{\color{blue}NOTE: This is a corrected version of the manuscript published as [Phys. Rev. A \textbf{80}, 011404(R) (2009)]. In the original manuscript, there was a sign error on the values of $\Delta\theta^{Ar}_{2q}$ used in the calibration procedure due to the different sign convention of Ref.\cite{Mairesse03} from which they were extracted. The given molecular phase differences were therefore too small by $2\Delta\theta_{2q}^\text{Ar}$, i.e., by $(0.4, 0.36,  0.32, 0.28)\:$rad for sidebands $(12, 14, 16, 18)$. The conclusions of our original paper, in particular the attribution of the molecular phases' behaviour to the presence of a resonance in the vicinity of harmonic 11 and their sensitivity to vibrational excitation of the ion, remain unaffected.}

We thank V. Blanchet, R. Cireasa, D. Dowek, A. Huetz, O. Lepelletier and H.J. W\"orner for fruitful discussions. We acknowledge financial support from the EU LASERLAB program and the French ANR program.

\end{document}